**An energy window study of light transmission-disorder relationship in 1D photonic structures**

Michele Bellingeri[1], Davide Cassi[1], Francesco Scotognella[2,3,*]

[1] Dipartimento di Scienze Matematiche, Fisiche e Informatiche, Università degli Studi di Parma, Parco Area delle Scienze 7/A, 43124, Parma, Italy
[2] Dipartimento di Fisica,, Politecnico di Milano, Piazza Leonardo da Vinci 32, 20133 Milano, Italy
[3] Center for Nano Science and Technology@PoliMi, Istituto Italiano di Tecnologia, Via Giovanni Pascoli, 70/3, 20133, Milan, Italy
* To whom the correspondence should be addressed: francesco.scotognella@polimi.it

**Abstract**
While the light transmission of photonic crystals is characterized by the photonic band gap, the one of disordered photonic structures is typified by a multiplicity of transmission depths. The total transmission over a range of wavelengths is related to the width of such range, but also to the type of disorder. Less homogeneous disordered structures transmit more light than the ordered counterpart regardless of the wavelengths range width. More homogeneous disordered structures transmit more light than the ordered counterpart only above a certain value of the width. We studied this behaviour with a statistical analysis over 5000 permutations of structures for each wavelength width and for each homogeneity degree (Shannon-Wiener index).

**Introduction**
The photonic crystals are characterized by the periodicity of refractive index materials presenting the most homogeneous structure. Altering the periodicity of the high and low refractive index materials we made photonic materials with a disorder level. The optical behaviour of a photonic crystal is characterized by an energy region in which light is not transmitted, the photonic band gap [1–6], and the energy regions that are transparent. In disordered photonic materials transmission valleys occur all over the studied energy range [7–12]. Thus, the light transmission, integrated in a certain energy interval, is related to the disorder of the structure. Indeed, the width of such energy interval is important in such comparison (e.g. within the photonic band gap region of the periodic photonic crystal). For example, it has been theoretically and experimentally demonstrated that disordered one-dimensional multi-layered photonic structures, in which the disorder arises via a random variation of the layer thicknesses, show a lower total transmission with respect to the corresponding periodic photonic crystal [13,14]. However, it has been demonstrated that the total transmission is also related, among the disordered photonic structures, to the homogeneity of the structures (in one dimension [15,16] and in two dimensions [17]). In the two-dimensional case, we observed a decrease of the total transmission by reducing the homogeneity, i.e., lower Shannon-Wiener index [17]. In the one-dimensional case, the total transmission increases by reducing the homogeneity; moreover, we observed that the total transmission for the most homogeneous disordered structures is lower with respect to the total transmission of the periodic photonic crystal [16].

In this work, we have studied the total light transmission through a periodic one-dimensional photonic crystal and disordered structures keeping constant the same number of high and low refractive index layers. For each degree of homogeneity (i.e. Shannon-Wiener index value) we compute 5000 different permutations of the layer sequence. We observed that less homogeneous disordered structures transmit more light than the ordered counterpart regardless of the wavelengths range width, while more homogeneous disordered structures transmit more light than the ordered counterpart only above a certain value of the width.

**Methods**

The periodic photonic crystal is composed by 16 unit cells, in which each unit cell include a 75 nm thick layer of $TiO_2$ and three 75 nm thick layers of $SiO_2$ (indeed, this is equivalent to a $SiO_2$ layer with a thickness of 225 nm). The disordered structures have been generated by the permutations of the sequence of the 64 layers and by measuring for each sequence the Shannon-Wiener index [18] of the sequence, given by the equation:

$$H' = \frac{-\sum_{j=1}^{S} p_j \log p_j}{\log(s)} \quad (1).$$

In the equation $p_j$ is a proportion of the $TiO_2$ layers in the *j*th unit cell and *s* is the number of unit cells in the photonic structure. Dividing by *log(s)* normalizes the index in the range (0,1). The Shannon index has been widely employed in diverse research areas, as information theory [19,20], statistics [21], statistical mechanics and physics [22], ecology [23–25], microbiology [26].

We chose two values of Shannon-Wiener index for the disordered photonic structures, corresponding to less homogeneous structures (H'=0.5507) and to more homogeneous structures (H'=0.9375). Thus, the script we wrote selected only the structures with these two values of Shannon-Wiener index (we report the script, written in MATLAB, in the Appendix).

To study the light transmission of the structure we assumed the incoming light to be at normal incidence. For a transverse electric (TE) wave the transfer matrix [16,27–29] for the *k*th layer is given by

$$M_k = \begin{bmatrix} \cos\left(\frac{2\pi}{\lambda} n_k d_k\right) & -\frac{i}{n_k} \sin\left(\frac{2\pi}{\lambda} n_k d_k\right) \\ -i n_k \sin\left(\frac{2\pi}{\lambda} n_k d_k\right) & \cos\left(\frac{2\pi}{\lambda} n_k d_k\right) \end{bmatrix} \quad (2)$$

with $n_k$ the refractive index and $d_k$ the thickness of the layer. The matrix for the multilayer (of *s* layers) is given by the product $M = M_1 \cdot M_2 \cdot \ldots \cdot M_k \cdot \ldots \cdot M_s = \begin{bmatrix} m_{11} & m_{12} \\ m_{21} & m_{22} \end{bmatrix}$. The transmission coefficient can be written as

$$t = \frac{2 n_s}{(m_{11} + m_{12} n_0) n_s + (m_{21} + m_{22} n_0)} \quad (3)$$

with $n_s$ the refractive index of the substrate (1.46) and $n_0$ the refractive index of air. Thus, the transmission is

$$T = \frac{n_0}{n_s} |t|^2 \quad (4).$$

The total transmission *tT* is given by the integral

$$tT = \int_{\lambda_C - HW}^{\lambda_C + HW} T(\lambda) \, d\lambda \quad (5)$$

where $\lambda_C$ is the wavelength corresponding to the photonic band gap centre and *HW* is the half width of the spectral window that we considered. The integral is performed with a wavelength step of 1 nm.

## Results and Discussion

The transmission spectrum of the periodic photonic crystal that we have selected is depicted in Figure 1a (black curve). For this structure $\lambda_c = 1054\ nm$ and we denoted this wavelength in Figure 1a (magenta line). The red and blue curves of Figure 1b indicate the light transmission of the disordered photonic structures with H'=0.5507 and H'=0.9375, respectively.

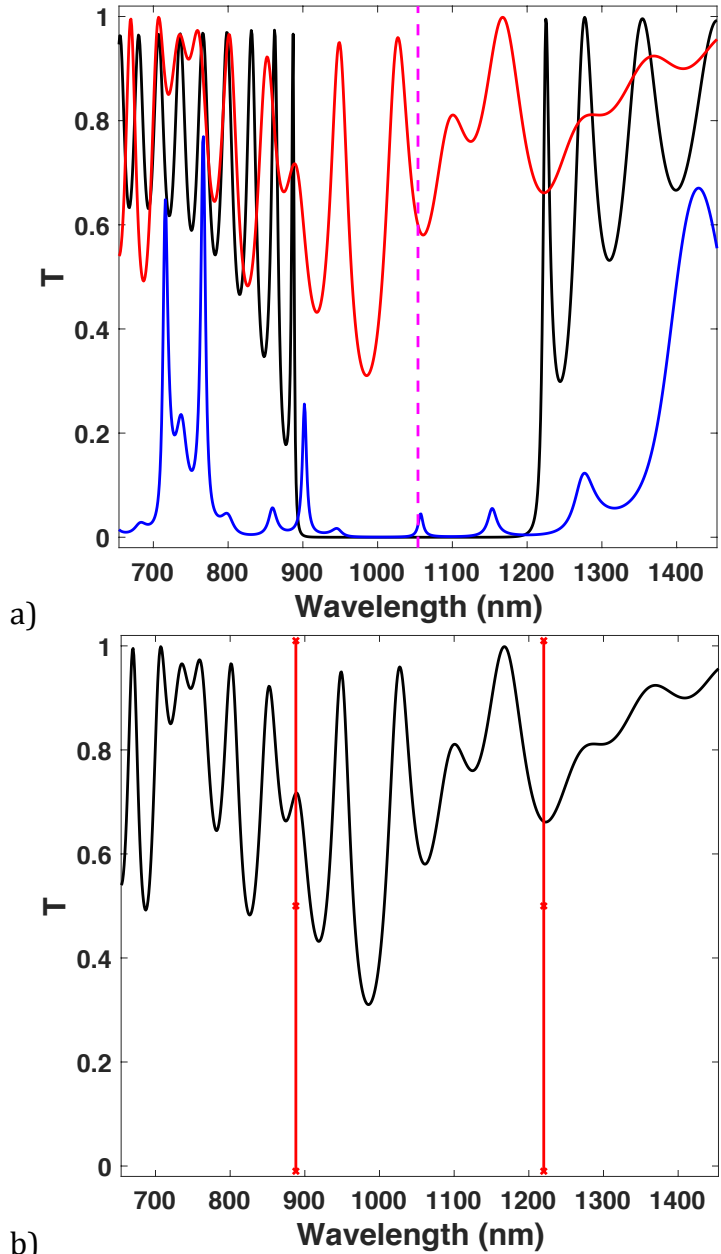

**Figure 1:** a) Light transmission spectrum of a 1D periodic photonic crystal (black curve), and disordered structures with H'=0.5507 (blue curve) and H'=0.9375 (red curve). The magenta line indicates $\lambda_c = 1054\ nm$. (b) Light transmission spectrum for the structure with H'=0.5507 (black curve) and the region in which the total transmission is integrated for $HW = 166\ nm$.

In Figure 1b the red lines denote the interval in which the total transmission is integrated, for the case of $HW = 166\ nm$. In this study we selected five values for $HW$: 166 nm (that corresponds to the half width at half maximum of the photonic band gap), 180 nm, 200 nm, 300 nm and 400 nm.

In Figure 2 we show the total transmission frequency distributions for the different window bandwidths around the band gap (*HW* values) for the photonic crystal with *H'*=0.5507. The vertical red line indicates the mean value of the total transmission. Range, mean values and related standard deviations are reported in Table 1. The total transmission frequency distributions for the photonic material *H'*=0.5507 are bell shaped with a slight left skewness, i.e. values lower than the mean are slightly more probable.

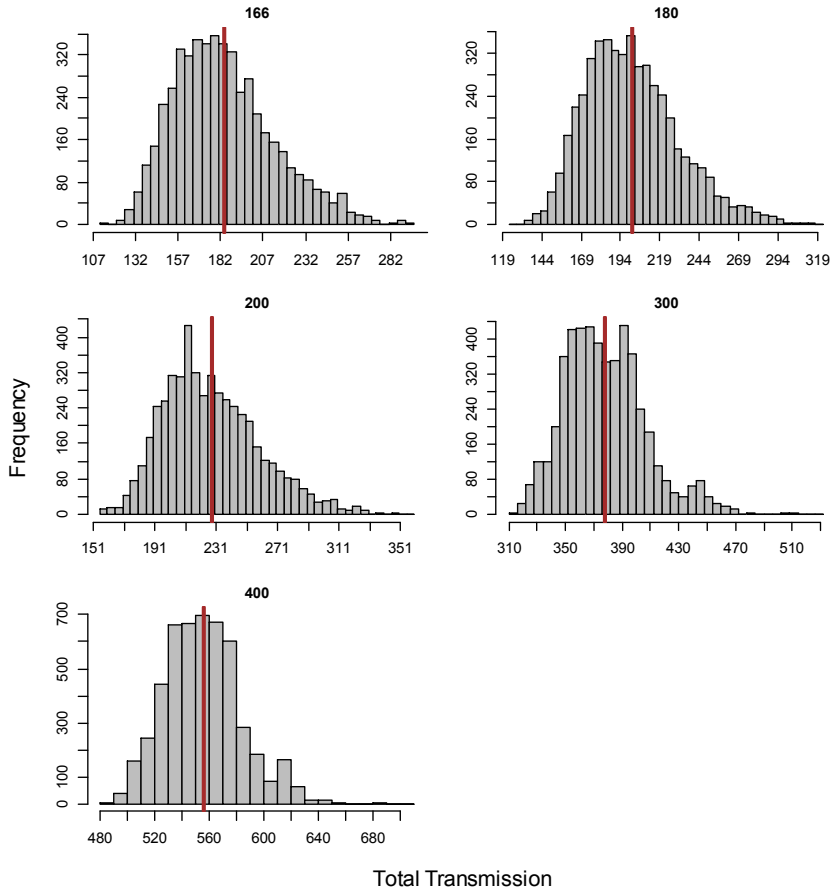

**Figure 2**: Total transmission frequency distributions as a function of the window bandwidth around the band gap (*HW*) for the photonic crystal with H'=0.5507. The vertical red line indicates the mean value of the total transmission.

**Table 1**: total transmission outcomes for the photonic crystal with H'=0.5507.

| range | µ | Σ | min | max |
|---|---|---|---|---|
| 166 | 184.0768 | 29.42569 | 112.4448 | 294.9343 |
| 180 | 201.2992 | 29.83337 | 124.2316 | 318.8463 |
| 200 | 227.6922 | 31.1228 | 156.1678 | 359.2722 |
| 300 | 378.068 | 28.71604 | 311.1484 | 531.5969 |
| 400 | 555.6224 | 27.98822 | 480.0845 | 705.2466 |

In Figure 3 we reported the total transmission frequency distributions as a function of the window bandwidths around the band gap for the photonic crystal with *H'*=0.9375. The vertical red line indicates the mean value of the total transmission. Range, mean values and related standard deviations are reported in Table 2. The total transmission frequency distributions for the photonic material *H'*=0.9375 show higher left skewness with a right tale than the *H'*=0.5507 photonic material, i.e. values of the total transmission higher than the mean are unlikely especially for the narrow bandwidths (166 and 180 nm, Figure 3, where we see a more pronounced right tale). Such left skewness could be ascribed to the fact that the lowest total transmission values are close to zero and thus the distribution is necessarily without the left tail.

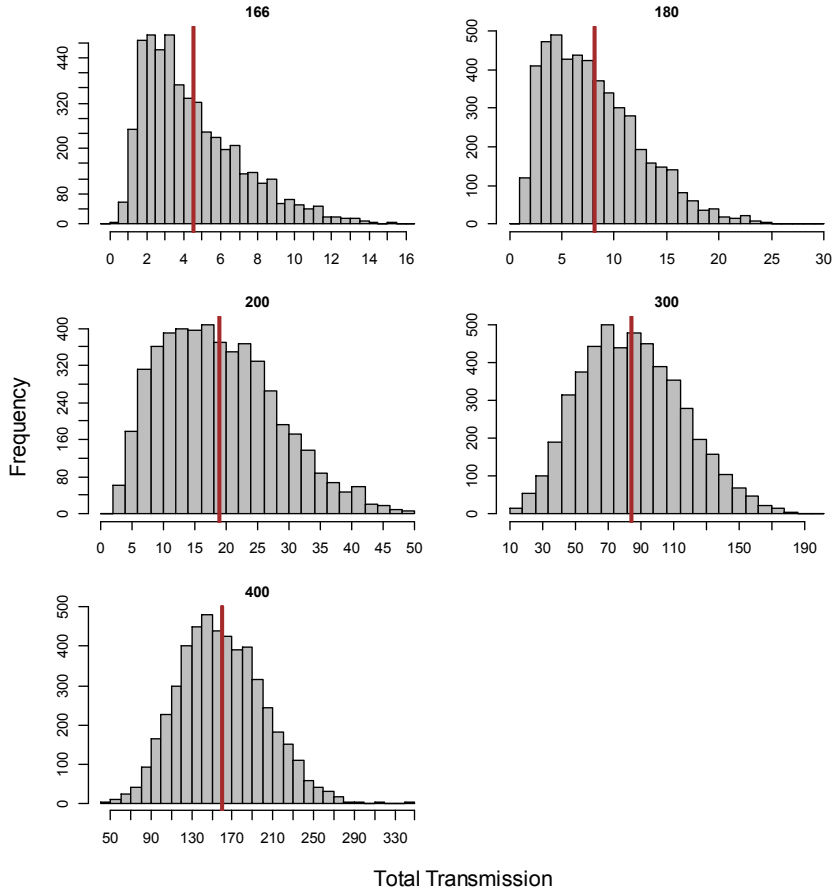

**Figure 3**: Total transmission frequency distributions as a function of the window bandwidth around the band gap (*HW*) for the photonic crystal with H'=0.9375. The vertical red line indicates the mean value of the total transmission.

**Table 2**: total transmission outcomes for the photonic crystal with H'=0.9375.

| range | μ | Σ | min | max |
|---|---|---|---|---|
| 166 | 4.493086 | 2.634135 | 0.4644 | 15.4964 |
| 180 | 8.072491 | 4.458453 | 0.9426 | 27.1352 |
| 200 | 18.91739 | 9.084675 | 1.9684 | 49.5562 |
| 300 | 84.68719 | 30.71156 | 12.5681 | 198.8361 |
| 400 | 159.6695 | 41.46815 | 42.351 | 349.9164 |

In Figure 4 we depict the total transmission as a function of the half width of the observed wavelength range *HW*. The circle blue curve represents the data of the structure with *H'*=0.5507; the diamond red line represents the data of the structure with *H'*=0.9375; the X sign black line represent the data of the periodic photonic crystal. Inset: magnification of the plot in the half width range 164 nm - 182 nm. In the magnification we highlight the crossover between the total transmission of the periodic photonic crystals and the disordered structure with *H'*=0.9375. Within the photonic band gap region ($HW = 166\,n$), the mean total transmission for the disordered structure with *H'*=0.9375 is higher than the one of the periodic photonic crystal. For larger width the disordered structure with *H'*=0.9375 transmits less light with respect to the periodic photonic crystal.

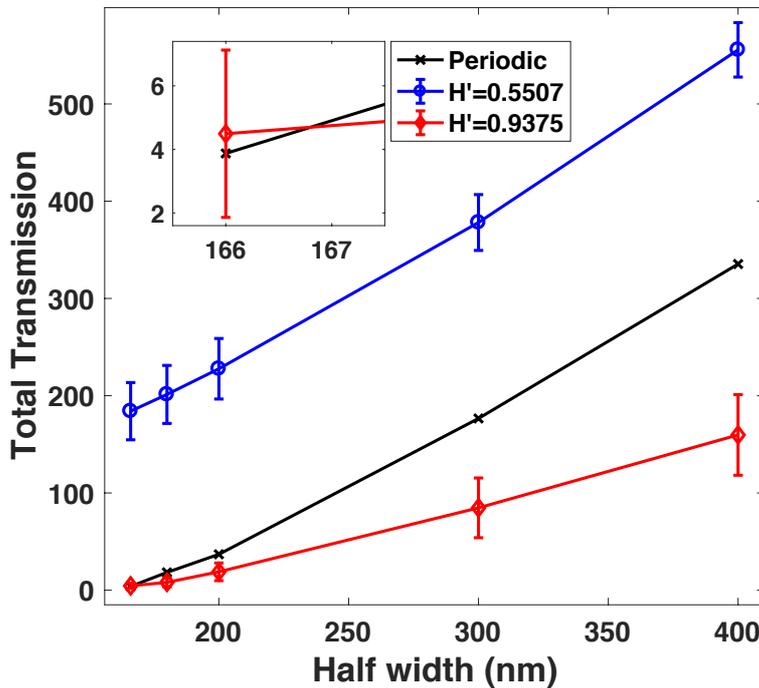

**Figure 4.** Total transmission as a function of the half width of the observed wavelength range. The circle blue curve represents the data of the structure with H'=0.5507; the diamond red line represents the data of the structure with H'=0.9375; the X sign black line represent the data of the periodic photonic crystal. Inset: magnification of the plot in the half width range 164 nm - 182 nm.

**Conclusion**
In this work, we have studied the total light transmission through a periodic one-dimensional photonic crystal and disordered structures keeping constant the same number of high and low refractive index layers and changing the homogeneity of the disordered structures. For each degree of homogeneity (i.e. Shannon-Wiener index value) we have computed 5000 different permutations of the layers sequence. We have observed that less homogeneous disordered structures transmit more light than the ordered counterpart regardless of the wavelengths range width, while more homogeneous disordered structures transmit more light than the ordered counterpart only above a certain value of the width.

**Acknowledgement**
This project has received funding from the European Union's Horizon 2020 research and innovation programme (SONAR) under the Marie Skłodowska-Curie grant agreement No. [734690].